\documentclass{iopjournal}

\usepackage{graphicx}
\usepackage{amsmath,amssymb}
\usepackage{bm}
\usepackage[numbers,sort&compress]{natbib}

\usepackage[utf8]{inputenc}
\usepackage[T1]{fontenc}
\usepackage{amsmath,amssymb}
\usepackage[T1]{fontenc}                  
\usepackage{algorithm}
\usepackage{algpseudocode}  
\begin{document}

	\articletype{Article type} %

	\title{Transducer leakage error suppression using invariant-based shortcut}

	\author{Shi-Qiang Qiao$^1$, Zheng Shan$^2$, Xue-Ke Song$^{1,*}$, Shi-Lei Su$^{3,4}$, Liu Ye$^1$ and Dong Wang$^{1,*}$}

	\affil{$^1$ School of Physics, Anhui University, Hefei 230601, China}
	\affil{$^2$ Laboratory for Advanced Computing and Intelligence Engineering, Zhengzhou 450001, China}
	\affil{$^3$ School of Physics, Zhengzhou University, Zhengzhou 450001, People's Republic of China}
	\affil{$^4$ Institute of Quantum Materials and Physics, Henan Academy of Science, Henan 450046, People's Republic of China}
	\affil{$^*$ Authors to whom any correspondence should be addressed.}

	\email{songxk@ahu.edu.cn, dwang@ahu.edu.cn}
	

	\begin{abstract}
  We present a method for suppressing transducer leakage errors in spin-superconducting hybrid quantum systems with the theory of optimal invariant-based shortcut. By mediated virtual photons as a transducer to exchange the energy between the spin qubit and a transmon qubit, the fidelity of the population of the final state features a broad range
  above 99\% under the influence of leakage error. The leakage probability from computational subspace to non-computational subspace can be effectively suppressed at a lowest value with $0.01$.
  Based on the optimized pulse control designed by the invariant-based inverse engineering, the high-fidelity quantum iSWAP gate operations and entanglement state preparation within the computational subspace are achieved.  {Compared to the traditional $\pi$ pulse, derivative removal by adiabatic gate,  counter-diabatic shortcut schemes, and limited-memory Broyden-Fletcher-Goldfarb-Shanno gradient ascent pulse engineering,  the optimized shortcut scheme can still achieve a high fidelity with 99\% in the presence of decoherence and control error.} When taking into account the possibility of leakage errors in actual situations, our solution can still largely resist the influence of control errors. The results provides a feasible path for precisely manipulating the quantum state of hybrid quantum systems.
	\end{abstract}

	\section{INTRODUCTION}
	
	The rapid development of quantum computing and quantum information processing has put forward higher requirements for the precise control of quantum systems. {Recently, the study of hybrid quantum systems has attracted considerable attention for their unique advantages of integrating different quantum platforms, which offer new possibilities for achieving high-performance and high-fidelity quantum state manipulation ~\cite{doi:10.1073/pnas.1419326112,RevModPhys.85.623,clerk2020hybrid,burkard2020superconductor}.}
 	{Among them, hybrid spin-superconducting systems are promising candidates for robust quantum information processing, in which semiconductor spin qubits such as resonant exchange (RX) type have a longer decoherence time~\cite{veldhorst2014addressable,yoneda2018quantum,PhysRevLett.121.177701,PhysRevLett.111.050501,PhysRevB.91.235411},} and superconducting qubits such as transmon type feature fast and high-fidelity gate operations and readout~\cite{barends2014superconducting,PhysRevApplied.10.054009,PhysRevA.110.022608,doi:10.1126/science.aaw1611,PhysRevApplied.16.064003}. The superconducting resonator acts as the mediation to achieve long-range coupling of spin
qubits and high-fidelity readout between the two qubits in both resonant and dispersive regimes~\cite{PhysRevX.7.011030,landig2019virtual,bottcher2022parametric}. It provides an attractive route to employ
the two-qubit interactions mediated by microwave photons to realize long-distance quantum state transfer and quantum gate operations in solid-state quantum systems.

    Traditional adiabatic methods can achieve high-fidelity quantum state manipulation in multiqubit quantum systems, while their slow operation speed makes them difficult to complete the required tasks within the decoherence time~\cite{RevModPhys.70.1003,RevModPhys.89.015006}. To address this issue, the shortcuts to adiabaticity (STA) technology emerged~\cite{TORRONTEGUI2013117,manzoni2017simulating}. STA can achieve the equivalent effects of the adiabatic evolution  in a short time through the design of optimized control pulses, thereby significantly increasing the operation speed and reducing error effects. The aim of STA is based on the principles of ``eliminating non-adiabatic transitions" and ``replicating adiabatic invariance". It can be achieved through methods such as counter-diabatic (CD) driving~\cite{demirplak2003adiabatic,Berry_2009,PhysRevLett.105.123003}  and Lewis-Riesenfeld (LR) invariants~\cite{10.1063/1.1664991,10.1063/1.525329,PhysRevA.95.022332}, and so on~\cite{UNANYAN199748,PhysRevLett.85.1626,Song_2016,Takahashi_2013,Zheng2022}. Furthermore, the STA technology can combine well with existing optimal control methods, such as dynamical decoupling, composite pulses, to improve the robustness the control of quantum systems
versus noise and perturbations or to optimize relevant variables. Thanks to the higher flexibility and outstanding compatibility of STA, it has been widely applied in multiple quantum research fields in recent years, such as quantum thermodynamics, non-Hermitian quantum systems, quantum metrology, and robust chiral discrimination~\cite{Song:21,doi:10.1021/acs.jpca.5b06090,Xu:23,math6070111,PhysRevA.93.052324,PhysRevLett.104.063002,PhysRevA.84.023415,d9yg-d3qr,Guo2025,Xing2026}.

{ Recently, the researchers experimentally observed the coherent interaction between the RX qubit and a transmon
qubit in the hybrid spin-superconducting systems, where virtual photon from superconducting resonator acts as a transducer to mediate the interaction coupling~\cite{landig2019virtual,scarlino2019coherent}.}  However, the complex interactions and multiple energy levels of such hybrid systems also brings new challenges, 
the leakage errors from the coupling between the high energy of superconducting qubits and the multi-electronic states of semiconductor quantum dots will affact the effectiveness of population swapping of quantum states or the population transitions 
 outside the computational subspace~\cite{PhysRevLett.111.050502,PhysRevA.76.042319,PhysRevA.111.012621,PhysRevLett.114.010501}. It is neccessary to design optimal control protocols to suppress the leakage errors, so that robust and high-fidelity transducer mechanism can be realized~\cite{Kim2025,Lin2026,lycw-g48q,Polat2026,Werninghaus2021,Guan2026,Chiaro2025}.
{A promising way is to employ the coherent control framework that combines the shortcuts to adiabaticity technology with the optimal control theory. }
Under this framework, a quantitative sensitivity index is introduced to analyze and characterize the sensitivity  to out-of-subspace transitions of the STA scheme~\cite{kiely2014inhibiting,PhysRevApplied.14.034003}.
The goal of the leakage suppression strategy is systematically optimizing the control pulses to minimize this sensitivity, thereby confining dynamics of the systems within the computational subspace, and finally a high-fidelity quantum state manipulation is achieved.
    
    In this paper, we propose an optimized pulse scheme based on STA to suppress the transducer leakage error in hybrid spin-superconducting quantum systems. By making use of the invariant-based shortcut and perturbation theory, we design the optimal control pulses to achieve the following targets: (1) effective suppression of leakage errors of transitions from computational subspace to non-computational subspace,  (2) { high-fidelity two-qubit iSWAP gate and entanglement generation within the computational subspace  $\{|00\rangle, |01\rangle, |10\rangle, |11\rangle\}$.}   
    Our numerical simulation results show that the optimized LR pulse scheme can suppress leakage probability from computational subspace to non-computational subspace at a lowest value with 0.01. {Compared to the traditional $\pi$ pulse, derivative removal by adiabatic gate (DRAG) pulse  , CD driving schemes, and limited-memory Broyden-Fletcher-Goldfarb-Shanno gradient ascent pulse engineering (LBFGS-GRAPE),} the optimizing protocol 
still achieve a fidelity of quantum state control with 99\%, showing its robustness against such errors. When taking into account the possibility of leakage error in actual situations, our solution can still largely resist the influence of control error. The findings offer an effective approach for mitigating leakage errors in high-dimensional hybrid quantum systems, demonstrating the broader applicability of optimal control theory in robust quantum computing and communication.
	
	The paper is organized as follows: Section ~\ref{sec2} introduces the theoretical model and the effective Hamiltonian of the hybrid quantum spin-superconducting system. Section ~\ref{sec3} present the leakage suppression by invariant-based optimal scheme. Section ~\ref{sec4} demonstrates the quantum gate operations and entanglement state preparation. Section ~\ref{sec5} gives a conclusion of the manuscript.
	\section{MODEL AND HAMILTONIAN}\label{sec2}

\begin{figure}
		\centering
		\includegraphics[width=10.6cm,height=5.86cm]{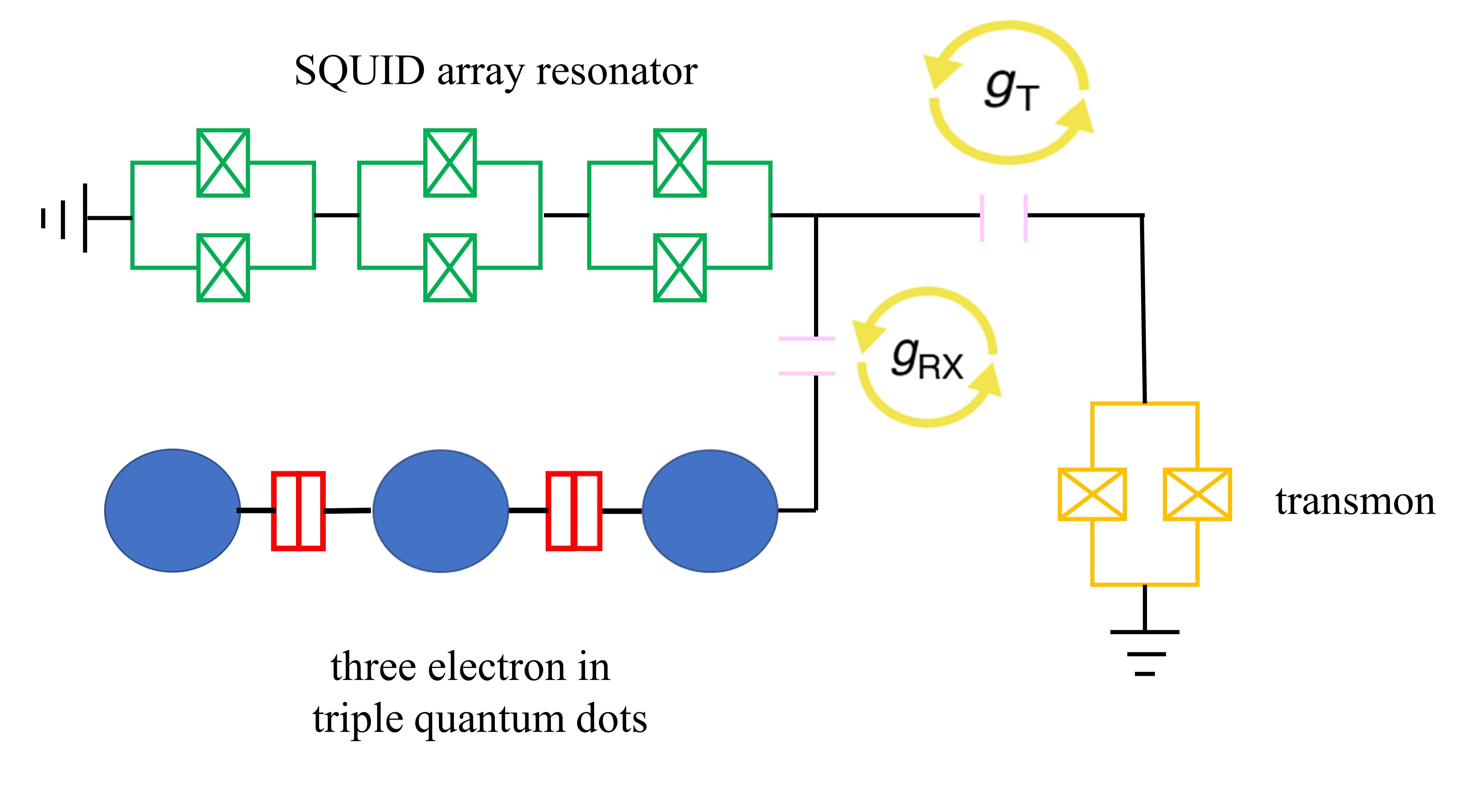}
		\caption{The mood of hybrid quantum system,which consists of an array of SQUID loops (green), an RX qubit (blue sphere), a transmon (orange). The yellow arrow indicates the coupling between quantum systems with a coupling strength of $g_{i}$. }\label{fig1}
	\end{figure}

	As illustrated in Fig.~\ref{fig1}, the hybrid quantum system consists of three different parts: a semiconductor spin qubit, a superconducting transmon qubit, and a superconducting resonator. {The RX type spin qubit consists of three electrons in a gallium arsenide (GaAs) triple quantum dot, which is capacitively coupled to the end of a superconducting quantum interference device (SQUID) array resonator with coupling strength $g_{RX}$. The superconducting transmon qubit consists of two Josephson junctions. The one end is connected to direct current (DC) ground, and the other end is capacitively coupled to the same end of a SQUID array resonator with the electric dipole coupling strength $g_{T}$.}
  The superconducting resonator is an  SQUID array loops, and its other end is connected to the DC ground.

	For SQUID array resonator, it is interacted with RX qubit and transmon qubit simultaneously. The Hamiltonian of resonator is
	\begin{eqnarray}\label{eq1}
		H_{r} = \omega_{r} a^{\dagger} a, 
	\end{eqnarray}
		where $\omega_{r}$ denote the frequency of resonator, $a^{\dagger}$ and $a$ are the creation and annihilation operations, respectively.
	
	For the RX qubit, three charge energy leves are defined as (1,1,1), (2,0,1), and (1,0,2), where $(n_l,n_m,n_r)$ indexes the number of electrons in left, middle, and right dots, respectively. Considering the subspace with total spin $S = \frac{1}{2}$ and spin z projection $S_z = \frac{1}{2}$, the basis vectors of this Hilbert subspace are~\cite{PhysRevLett.111.050501,PhysRevB.91.235411}  
	\begin{eqnarray}\label{eq2}
		\cr\cr|0\rangle&=&\frac{1}{\sqrt{2}}(|\uparrow, \uparrow, \downarrow\rangle-|\downarrow, \uparrow, \uparrow\rangle)		
		\cr\cr|1\rangle&=&\frac{1}{\sqrt{6}}(2|\uparrow, \downarrow, \uparrow\rangle-|\uparrow, \uparrow, \downarrow\rangle-|\downarrow, \uparrow, \uparrow\rangle)
		\cr\cr|2\rangle&=&\frac{1}{\sqrt{2}}\left(|\uparrow \downarrow\rangle_{l}-|\downarrow \uparrow\rangle_{l}\right)|\cdot\rangle_{m}|\uparrow\rangle_{r}
		\cr\cr|3\rangle&=&\frac{1}{\sqrt{2}}|\uparrow\rangle_{l}|\cdot\rangle_{m}\left(|\uparrow \downarrow\rangle_{r}-|\downarrow \uparrow\rangle_{r}\right). 
	\end{eqnarray}
    where  $|\uparrow\rangle$ and $|\downarrow\rangle$ represent the electron spin state, and $|\cdot\rangle$ denotes the vacuum state. Here, the state $|0\rangle$ and $|1\rangle$ are corresponding charge configuration of three quantum dots with (1,1,1), $|2\rangle$ and $|3\rangle$ denote the charge configurations of three quantum dots with (2,0,1) and (1,0,2), respectively.
    {In the relevant subspace $\{{|0\rangle, |1\rangle, |2\rangle, |3\rangle}\}$, the Hamiltonian of the quantum dot system can be expressed as}
    \begin{eqnarray}\label{eq3}
    	H_{s}^{0}=\begin{pmatrix} 
    		0 & 0 & \frac{t_{l}}{2} & \frac{t_{r}}{2} \\ 
    		0 & 0 & \frac{\sqrt{3} t_{l}}{2} & \frac{\sqrt{3} t_{r}}{2} \\ 
    		\frac{t_{l}}{2} & \frac{\sqrt{3} t_{l}}{2} & -\epsilon-\delta & 0 \\ 
    		\frac{t_{r}}{2} & \frac{\sqrt{3} t_{r}}{2} & 0 & \epsilon-\delta
    	\end{pmatrix},
    \end{eqnarray}
      where $t_{r} (t_{l})$ denotes the amplitude of coupling strength between the right (left) quantum dot and the middle quantum dot. The parameters $\epsilon$ and $\delta$ indicate $(E(1,0,2)-E(2,0,1))/2$ and $E(1,1,1)-(E(1,0,2))+E(2,0,1))/2$, respectively, where $E(a,b,c)$ denote energy of the $(a,b,c)$. By diagonalizing Eq.~(\ref{eq3}), we can rewrite the Hamiltonian of the RX qubit as
    \begin{eqnarray}\label{eq4}
    	H_s = \sum_{j=0}^{3} \omega_{s,j}|j\rangle\langle j|.
    \end{eqnarray}
    where $\omega_{s,j}$ denote the frequency of $|j\rangle$ $(j=0,1,2,3)$ state of the spin qubit with $\omega_{s,0}=0$. Moreover, we can encode the spin qubits by using the lowest two energy level of RX qubit, $|0_{RX}\rangle$ and  $|1_{RX}\rangle$. In this case, the ground and excited states of RX qubit can be written as linear combinations of spin states $|0\rangle, |1\rangle, |2\rangle$, and $|3\rangle$. This gives
    \begin{eqnarray}\label{eq5}
    |0/1_{RX}\rangle=\sum_{j=0}^{3}c_{j}^{(0/1)}|j\rangle.
    \end{eqnarray}
    
    Under the rotating wave approximation, the Hamiltonian describing the interaction of RX qubit and resonator can be expressed as~\cite {landig2018coherent}
    \begin{eqnarray}\label{eq6}
     H_{s,r}=g_{s,j}(a^{\dagger}\sigma_{j-1,j}+a\sigma_{j,j-1}),
    \end{eqnarray}
where  $\sigma_{j-1,j}=|j-1\rangle\langle j|$ and the $g_{s,j}$ is the coupling strength between the RX qubit and the resonator with
\begin{eqnarray}
     g_{s,j}=2 g_{RX}\left[c_{2}^{1} c_{2}^{0_{*}}-\alpha\left(c_{3}^{1} c_{3}^{0_{*}}+c_{2}^{1} c_{2}^{0_{*}}\right)\right],
    \end{eqnarray}
      with $g_{RX}$ being the charge-photon coupling strenth. The coefficients $c_{i}^{0}$ and $c_{i}^{1}$ are defined from Eq.~(\ref{eq5}).
    
      For transmon, it can be seen as a multilevel superconducting qubit, whose Hamiltonian reads
    \begin{eqnarray}\label{eq7}
    	H_{t}=\sum_{k=0}^{N}\omega_{t,k}\sigma_{k,k},
    \end{eqnarray}
   where $\omega_{t,k}$ is frequency of transmon state $|k\rangle$, and $\omega_t = \sqrt{8 E_{C} E_{J}} - E_{C}$ with $E_{C}$ and $E_{J}$ representing energy of charging and Josephson junction. The operator $\sigma_{k,k}=|k\rangle\langle k|$. Here, we take the number of transmon levels of $N=3$ as a cutoff.
	
 Similarly, the Hamiltonian for the interaction of the transmon and the resonator takes the form of~\cite{PhysRevA.75.032329,slichter2016quantum}
	\begin{equation}\label{eq8}
		H_{t,r}=\sum_{k=1}^{3} g_{t, k}(a^{\dagger}\sigma_{k-1,k}+ a\sigma_{k, k-1}), 
	\end{equation}
   where $g_{t,k}$ is coupling strength between the transmon and the superconducting resonator. 
 
        As a result, the total Hamiltonian of the full system consisting of the transmon, RX qubit, and the resonators can be described by
        \begin{equation}\label{eq9}
        \begin{aligned}
        	H_{\text{tot}} &= H_{s} + H_{t} + H_{r} + H_{s,r} + H_{t,r} \\
        	&= \sum_{j=1}^{3} \omega_{s, j}|j\rangle\langle j| + \sum_{k=1}^{3} \omega_{t, k}|k\rangle\langle k| + \omega_{r} a^{\dagger} a \\
        	& + \sum_{j=1}^{3} g_{s, j}\left(a^{\dagger} \sigma_{s, j}^{-} + a \sigma_{s, j}^{+}\right) + \sum_{k=1}^{3} g_{t, k}\left(a^{\dagger} \sigma_{t, k}^{-} + a \sigma_{t, k}^{+}\right), 
        	          \end{aligned}
       \end{equation}
      where $\sigma^{+}_{s,j}=|j\rangle\langle j-1|$ , $\sigma^{-}_{s,j}=|j-1\rangle\langle j|$, and  $\sigma^{+}_{t,k}=|k\rangle\langle k-1|$ , $\sigma^{-}_{t,k}=|k-1\rangle\langle k|$.
      
        We consider the whole system in the dispersive regime, where the frequencies
       of both qubits are detuned from the resonator. For example, typical coupling strength between the RX (transmon) qubit and the resonator are approximately 90(200) MHz, with the qubit-resonator detuning set to 2 GHz~\cite{landig2019virtual}. In this case, the effective coupling between the RX qubit and transmon qubit is mediated by virtual photons in resonator.
The effective Hamiltonian by using Schrieffer-Wolff transformations~\cite{PhysRev.149.491} decouple different subspaces can be obtained as 
        \begin{eqnarray}\label{eq10}
       \begin{aligned}
       	H_{\text{eff}} &= \sum_{j=1}^{3} \omega_{s, j}'|j\rangle\langle j| + \sum_{k=1}^{3} \omega_{t, k}'|k\rangle\langle k| \\
       	&  + \sum_{j=1}^{3} \sum_{k=1}^{3} J_{j, k}\left(\sigma_{s, j}^{+} \sigma_{t, k}^{-} + \sigma_{s, j}^{-} \sigma_{t, k}^{+}\right),
       \end{aligned}
       	\end{eqnarray}
     where $\omega_{s,j}^{\prime}=\omega_{s,j}-\frac{g_{s,j}^{2}}{\Delta_{s,j}}$, $\omega_{t,j}^{\prime}=\omega_{t,j}-\frac{g_{t,j}^{2}}{\Delta_{t,j}}$, and the effective coupling strength of RX qubit and transmon qubit being $J_{j,k}=\frac{g_{s,j}g_{t,k}}{2}(\frac{1}{\Delta_{s,j}}+\frac{1}{\Delta_{t,k}}).$ The dynamics of the hybrid spin-transmon system are governed by the effective Hamiltonian $H_{\text{eff}}$.
   
    Here, we focus on suppressing transducer leakage errors and implementing the iSWAP gate of this hybrid quantum system in the computational subspace, spanned by the basis $\{|00\rangle , |01\rangle , |10\rangle , |11\rangle\}$. The remaining higher-energy states introduced by the multi-electron nature of the RX qubit and the transmon's anharmonicity act as detrimental leakage channels which must be suppressed to achieve high-fidelity operations.  { To this end, we employ the invariant-based shortcut approach to provide a systematic method for actively minimizing leakage within the framework of perturbation theory.}  With leakage error suppressed, this control framework can be applied to realize high-performance quantum operations within the computing subspace, including rapid and high-fidelity state transfer, the robust implementation of  the iSWAP gate, and the preparation of entangled states.    
    \section{OPTIMAL CONTROL DESIGN AND LEAKAGE SUPPRESSION}\label{sec3}
    In this section, we design driving pulses using optimal shortcuts to adiabaticity to realize a perfect population transfer of hybrid systems, and then show that the leakage error can be effectively suppressed in the case of the existence of uncontrollable transitions.
    \subsection{ Hamiltonian Decomposition and Leakage Model }\label{sec3.1}
 For the system Hamiltonian in Eq.~(\ref{eq10}), we first decompose the system Hamiltonian into the ideal control Hamiltonian  $H_{\text{sub}}$ and the leakage error Hamiltonian $H_{\text{leak}}$, which are used to describe the manipulation of quantum state within the calculation subspace and the leakage effect outside the subspace. {Subsequently, one can obtain the interaction Hamiltonian between spin qubits and superconducting qubits to govern the evolution of the system, in the basis $\{|00\rangle , |01\rangle , |10\rangle , |11\rangle\}$, by applying two unitary transformations on the system Hamiltonian~\cite{PhysRevA.111.012621}.} At this point the interaction Hamiltonian can be rewritten as
    \begin{equation}\label{eq11}
   H_{\text{int}} = H_{\text{sub}} + H_{\text{leak}}.
    \end{equation}
  with
    \begin{equation}\label{eq12}
    \begin{aligned}
    H_{\text{sub}} = \frac{\Delta}{2}\left(|10\rangle\langle 10| - |01\rangle\langle 01|\right) + \frac{\Omega}{2}\left(|01\rangle\langle 10| + |10\rangle\langle 01|\right),
    \end{aligned}
    \end{equation}
   and
     \begin{equation}\label{eq13}
  \begin{aligned}
  	H_{\text{leak}} &= \left(\frac{\Delta}{2} + \delta \omega_{s, 21}'\right)|20\rangle\langle 20| + \left(-\frac{\Delta}{2} + \delta \omega_{t, 21}'\right)|02\rangle\langle 02| \\ 
  	& + \frac{\eta_{0102} \Omega}{2}\left(|01\rangle\langle 02| + |02\rangle\langle 01|\right) \\
  	& + \frac{\eta_{1020} \Omega}{2}\left(|10\rangle\langle 20| + |20\rangle\langle 10|\right) \\
  	& + \frac{\eta_{0203} \Omega}{2}\left(|02\rangle\langle 03| + |03\rangle\langle 02|\right)\\
  	& + \frac{\eta_{2030} \Omega}{2}\left(|20\rangle\langle 30| + |20\rangle\langle 30|\right),
  \end{aligned}
   \end{equation} 
   {where $\Delta=d[(\omega_{t,1}'-\omega_{s,1}')]/dt$ is frequency modulation dependent time,  $\Omega=2J_{1,1}$ denotes exchange coupling between state $|01\rangle$ and $|10\rangle$, and $\eta$ is coefficient of the coupling strength in states relative to $\Omega$.} $\delta\omega'_{s,jj}$ is the frequency difference between the states $|j\rangle$ and the $|j\rangle$ of spin qubit, and $\delta\omega'_{t,kk}$ is the frequency difference between the state  $|k\rangle$ and the  $|k\rangle$ of transport qubit. 
   {In order to simplify the calculation, we set $\eta_{0102}=\eta_{1020}=\eta_{0203}=\eta_{2030}=\eta$, $\delta\omega'_{s,jj}=\delta\omega'_{t,kk}=\delta\omega_{kk}$.} Here, the  basis vectors \{$|03\rangle,|30\rangle$\} that are not directly coupled with \{$|01\rangle,|10\rangle$\} can be ignored since the suppressions of the transitions between $|01\rangle$ ($|10\rangle$) and $|02\rangle$ ($|20\rangle$) will lead to a very small value of the transition between $|02\rangle$ ($|20\rangle$) and  $|03\rangle$ ($|30\rangle$), and thus these two transitions can be disregarded. Finally, within the subspace spanned by the basis vectors $\{|01\rangle , |02\rangle , |10\rangle , |20\rangle\}$, the Hamiltonian of the system can be represented as
     \begin{equation}\label{eq15}
    \begin{gathered}
    	H_{\mathrm{sys}}=\begin{pmatrix}-\frac{\Delta}{2}&\frac{\eta\Omega}{2}&\frac{\Omega}{2}&0\\\frac{\eta\Omega}{2}&-\frac{\Delta}{2}+\delta\omega_{21}&0&\frac{\eta\Omega}{2}\\\frac{\Omega}{2}&0&\frac{\Delta}{2}&\frac{\eta\Omega}{2}\\0&\frac{\eta\Omega}{2}&\frac{\eta\Omega}{2}&\frac{\Delta}{2}+\delta\omega_{21}
    	\end{pmatrix}\end{gathered}	.
    \end{equation}
    \subsection{Invariant-Based Shortcut for Error Sensitivity Optimization}\label{sec3.2}
		To achieve high-fidelity quantum control of quantum state in the computational subspace, we employ the LR invariant shortcut  to design the optimal control pulses.
			 Considering the leakage space outside the computational subspace, 
		we treat the leakage error as a perturbation to the ideal subspace dynamics. We can rewrite Hamiltonian in Eq.~(\ref{eq15}) as
		\begin{equation}\label{eq21}
			H_{\text{sys}}=H_{0}+\eta V(t),
		\end{equation}
		where $H_0=H_{\text{sys}}|_{\eta=0}$,
		 \begin{equation}\label{eq41}
			\begin{gathered}
				H_{0}=\begin{pmatrix}-\frac{\Delta}{2}&0&\frac{\Omega}{2}&0\\0&-\frac{\Delta}{2}+\delta\omega_{21}&0&0\\\frac{\Omega}{2}&0&\frac{\Delta}{2}&0\\0&0&0&\frac{\Delta}{2}+\delta\omega_{21}
			\end{pmatrix}\end{gathered}	,
		\end{equation}
		 and 
		\begin{equation}\label{eq22}
			V=\begin{pmatrix}0&\frac{\Omega}{2}&0&0\\\frac{\Omega}{2}&0&0&\frac{\Omega}{2}\\0&0&0&\frac{\Omega}{2}\\0&\frac{\Omega}{2}&\frac{\Omega}{2}&0\end{pmatrix}.
		\end{equation} 
		We can find that the basis $|01\rangle $ is coupled with the basis $|10\rangle$, and both of them are decoupled from the subspace $|02\rangle$ and $|20\rangle$. In this case, studying the dynamics governed by the Hamiltonian  $H_0$ can be reduced to manipulate the time evolution of a two-level Hamiltonian $H_{\rm sub}$ in the subspace of $\{|01\rangle, |10\rangle\}$. 
Based on the Lewis-Riesenfeld theory, the dynamical invariant satisfying the equation $\frac{\partial}{\partial t}I+\frac{\mathrm{i}}{\hbar}[H_{\rm sub},I]=0$~\cite{Ruschhaupt_2012} can be expressed as 
		\begin{eqnarray}
		I(t)=\frac{1}{2}\mu{\begin{pmatrix}\cos\theta&\sin\theta e^{-i\beta}\\\sin\theta e^{i\beta}&-\cos\theta\end{pmatrix}},
		\end{eqnarray}
		 where $\mu$ is an arbitrary constant with units of frequency to
		keep $I(t)$ with dimensions of energy, and the time-dependent control parameters $\theta$ and $\beta$, by solving the above dynamical equation, can be given by
		\begin{eqnarray}\label{eq16}
			\dot{\theta}&=&-\Omega\sin\beta,\cr\cr
			\dot{\beta}&=&-\Delta-\Omega\cot\theta\cos\beta.
		\end{eqnarray}
	The eigenstates of invariant $I(t)$ are
		\begin{equation}\label{eq17}
			|\phi_+(t)\rangle=\begin{pmatrix}\cos\frac{\theta}{2}e^{-i\beta}\\\sin\frac{\theta}{2}\end{pmatrix},\quad
			|\phi_-(t)\rangle=\begin{pmatrix}\sin\frac{\theta}{2}\\-\cos\frac{\theta}{2}e^{i\beta}\end{pmatrix},
		\end{equation}
		with eigenvalues $\pm\frac{1}{2}\mu$. Based on LR theory, the general solution of the Schrödinger equation can be written as the superposition of the two orthogonal dynamical modes of the
		invariant $I(t)$ as
		\begin{equation}
			|\Psi(t)\rangle=\Sigma_nc_n|\phi_n(t)\rangle e^{i\gamma_n},
			\end{equation}
			where $c_n$ is a time-independent constant and the $\gamma_n$ are LR phases 
		\begin{equation}
			\gamma_{\pm}=\pm\frac{1}{2}\int_{0}^{t}dt^{\prime}\left(\dot{\beta}+\frac{\dot{\theta}\cot\beta}{\sin\theta}\right).
		\end{equation}
		 Inspired by the above method, when all the subspace in the Hamiltonian $H_0$ are considered,
	 the time-dependent Schrödinger solution for the this Hamiltonian can be expressed as follows

		\begin{eqnarray}\label{eq25}
			\begin{aligned}
				&|\phi_0(t)\rangle=\begin{pmatrix}\cos\frac{\theta}{2}e^{-i\beta}\\0\\\sin\frac{\theta}{2}\\0\end{pmatrix}e^{i\gamma_{+}},\\
				&|\phi_1(t)\rangle=\begin{pmatrix}\sin\frac{\theta}{2}\\0\\-\cos\frac{\theta}{2}e^{i\beta}\\0\end{pmatrix}e^{i\gamma_{-}},\\
				&|\phi_2(t)\rangle=\begin{pmatrix}0\\e^{-\frac{i}{2}\Gamma_1}\\0\\0\end{pmatrix},\\
				&|\phi_3(t)\rangle=\begin{pmatrix}0\\0\\0\\e^{-\frac{i}{2}\Gamma_2}\end{pmatrix},
			\end{aligned}
		\end{eqnarray}	 
			 {where $\dot\Gamma_1=\left(-\frac{\Delta}{2}+\delta\omega_{21}\right)$ and $\dot\Gamma_2=\left(\frac{\Delta}{2}+\delta\omega_{21}\right)$}.
		Our objective is to design an adiabatic shortcut based on LR invariants to achieve a particle number inversion from state $|01\rangle$ to state $|10\rangle$  along the invariant eigenstate $|\phi_0(t)\rangle$ within a given time $T$, with the overall phase factor being ignored. To ensure that the desired initial and final states can be obtained, we first provide the boundary conditions that the paramter $\theta$ needs to satisfy:
				\begin{equation}\label{eq40}
			\theta(0) = 0, \quad \theta(T) = \pi.
			\end{equation}
			In addition, one should guarantee that the Hamiltonian $H_{\rm sub}(t)$ and invariants $I(t)$ take the common eigenstate in the initial and final time, that is, $[H_{\rm sub}(0),I(0)]=0$ and  $[H_{\rm sub}(T),I(T)]=0$ are also required. Additional boundary conditions also need to be met
		\begin{eqnarray}\label{eq42}
				\begin{aligned}
			&\Omega(0) = 0, &\quad \dot{\theta}(0) = 0,\\
			&\Omega(T) = 0, &\quad \dot{\theta}(T) = 0. 
		\end{aligned}
			\end{eqnarray}
		By using time-dependent perturbation theory, we can calculate the fidelity to be found at the final state $|10\rangle$ from initial state $|01\rangle$
		\begin{equation}\label{eq23}
			F=1-\eta^2q+\mathcal{O}(\eta^4),
			\end{equation}
		where error sensitivity is 
		\begin{equation}\label{eq24}
			q=\frac{1}{2}\sum_{k=0}^3\left|\int_0^T\mathrm{d}t\left\langle\psi_0(t)|V(t)|\psi_k(t)\right\rangle\right|^2.
			\end{equation}

	Substitute Eq.~(\ref{eq25})  into  Eq.~(\ref{eq24}), we can obtain the analytic expression of the error sensitivity as
	\begin{equation}\label{eq26}
		\begin{aligned}
	 q&=\frac{1}{8}\left|\int_0^Tdt\left(\sin\frac{\theta}{2}\Omega e^{-\frac{i}{2}\Gamma_1-i\gamma_{+}}\right)\right|^{2} 
			+\frac{1}{8}\left|\int_0^Tdt\left(\cos\frac{\theta}{2}\Omega e^{-\frac{i}{2}\Gamma_2-i\gamma_{+}+i\beta}\right)\right|^{2}.
		\end{aligned}
		\end{equation}
	Following the idea of Refs.~\cite{Ruschhaupt_2012,PhysRevLett.111.050404,PhysRevA.88.033406}, when using Fourier series type of Ansatz 
	\begin{equation}\label{eq27}
		2\gamma_{+}-\beta=\nu(2\theta-\sin(2\theta)),
		\end{equation}
       we can get 
	\begin{equation}\label{eq28}
		\beta=\operatorname{arccot}(4\nu\sin^3\theta).
		\end{equation}
				\begin{figure}[h]
			\centering
			\includegraphics[width=5.6cm,height=4.6cm]{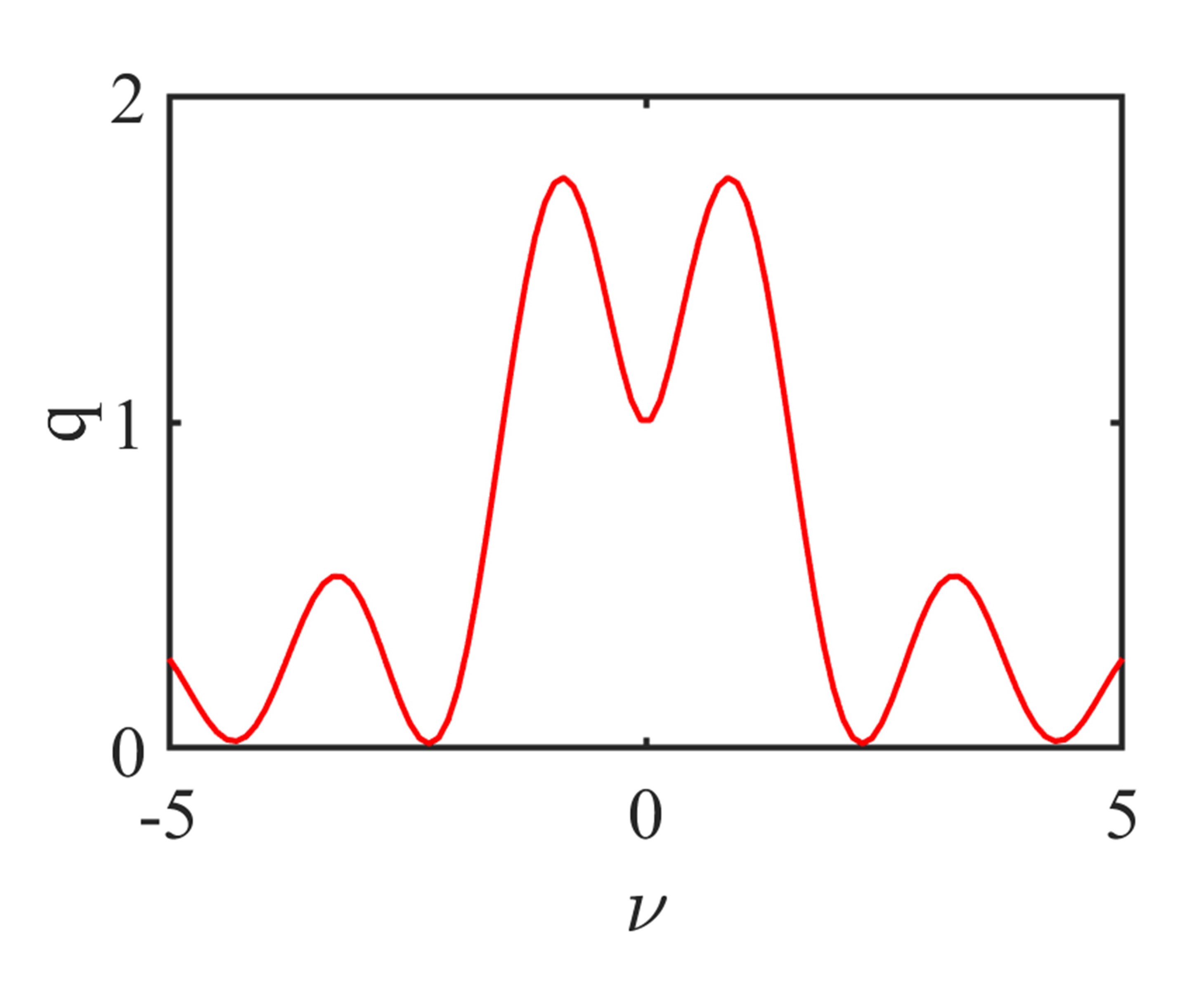}
			\caption{ Leakage error sensitivity q in Eq.~(\ref{eq26}). For specific
				values of $|\nu|$, $q=0$ is satisfied, in particular at the minimal value
				$|\nu|$=2.27.}\label{fig7}
		\end{figure}	
		\begin{figure}[h]
			\centering
			\includegraphics[width=10.6cm,height=5.2cm]{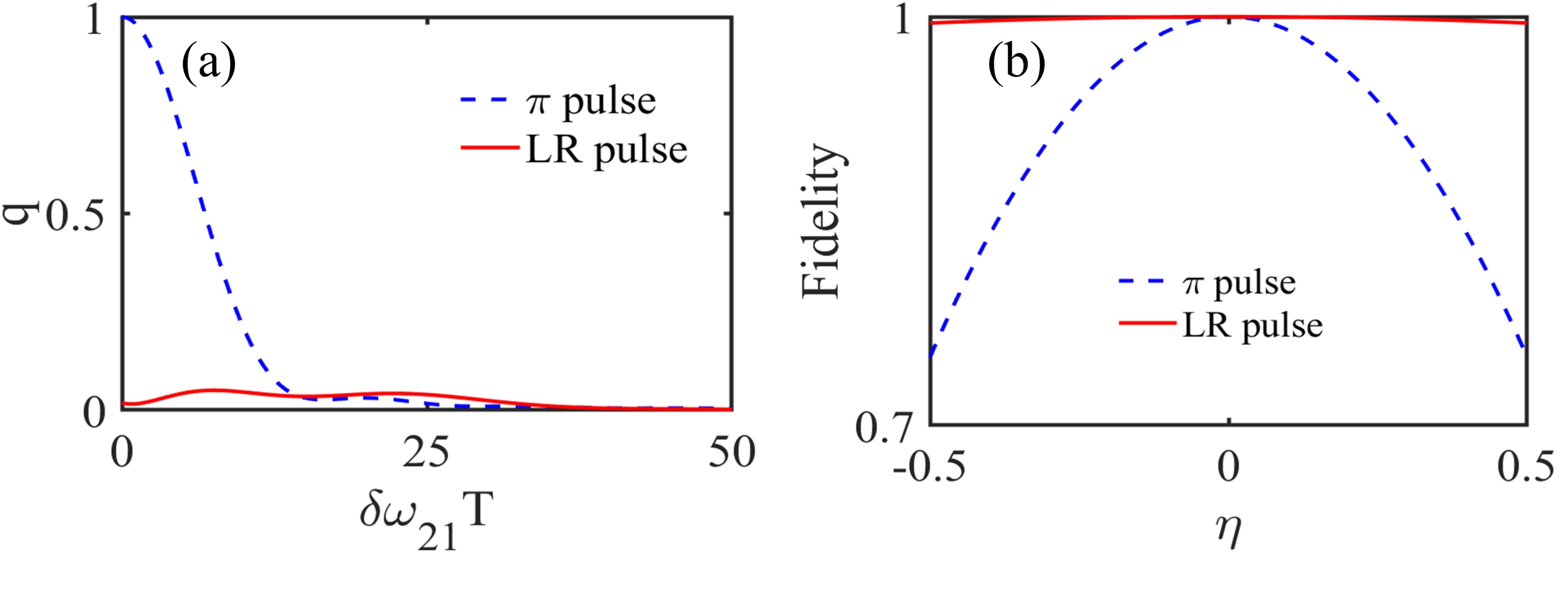}
			\caption{(a) The variation of error sensitivity with $\delta\omega_{21}$, which is the frequency difference between
				the  state $|2\rangle$ and state $|1\rangle$ for $\pi$ pulse scheme (blue, dotted line) and LR optimized pulse scheme (red, solid line) given by solving Eq.~(\ref{eq16}). (b) We obtain schematic diagram of fidelity by Eq.~(\ref{eq23}), with make $\delta\omega_{21}=0$  for $\pi$ pulse scheme (blue, dotted line) and LR optimized pulse scheme (red, solid line).}\label{fig3}
		\end{figure}
	Inversely, from Eq.~(\ref{eq16}), the corresponding Rabi frequency $\Omega$ and the detuning $\Delta$ of the Hamiltonian can be calculated as 
	\begin{eqnarray}\label{hp}
		\Omega&=&-\dot{\theta}\sqrt{1+16\nu^2\sin^6\theta},\cr\cr
		\Delta&=&16\nu\dot{\theta}\sin^{2}\theta\cos\theta\frac{1+4\nu^{2}\sin^{6}\theta}{1+16\nu^{2}\sin^{6}\theta}.
		\end{eqnarray}
	
	To achieve the goal of population transfer from $|01\rangle$ to $|10\rangle $, we choose a smooth function of $\theta$ to satisfy the boundary conditions as
	\begin{equation}\label{eq29}
		\theta=\frac{\pi}{2}\left[1+\sin\frac{\pi(2t-T)}{2T}\right].
	\end{equation}
	Also, the relationship between error sensitivity $q$ and $\nu$ can be obtained, which is shown in Fig.~\ref{fig7}. To minimize the $q$, we select the value of $\nu$ as 2.27.

			\begin{figure}
			\centering
			\includegraphics[width=10.6cm,height=5.6cm]{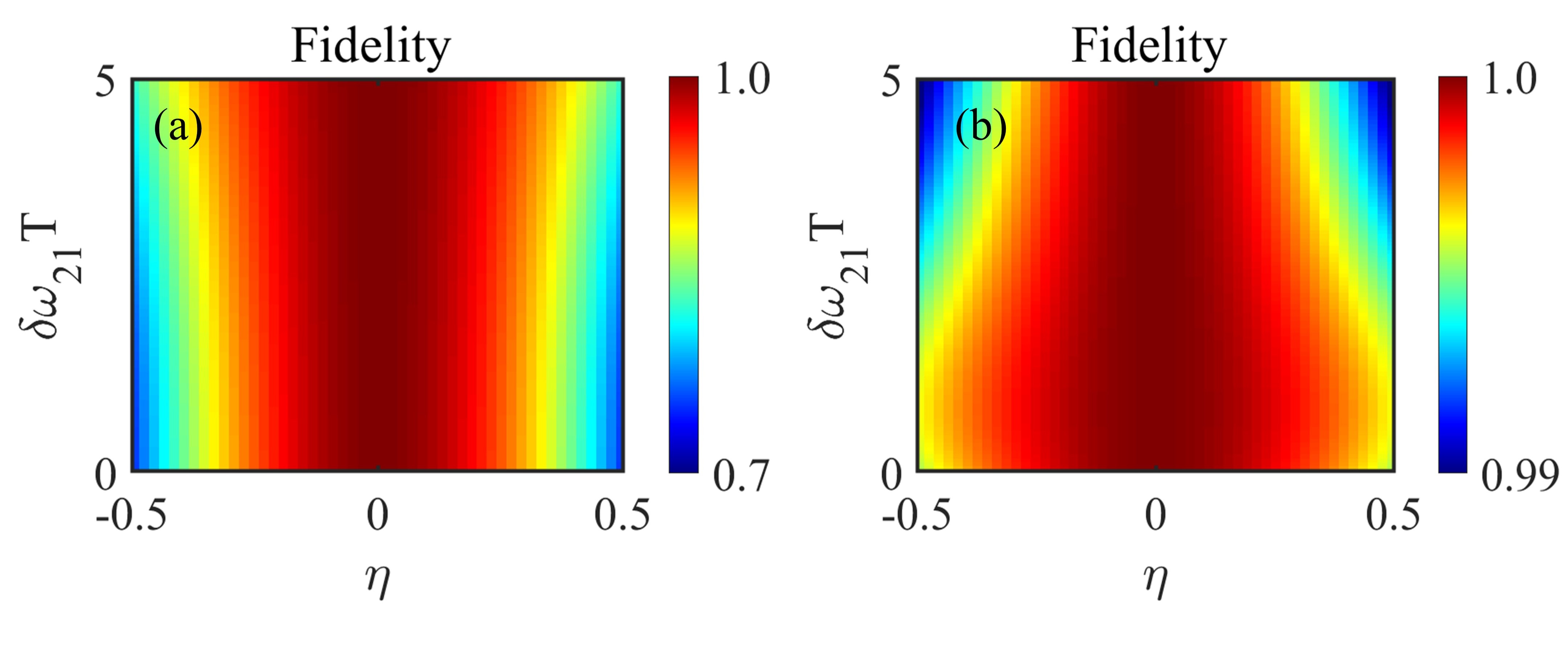}
			\caption{ The variations of fidelities with respect to the parameters $\delta\omega_{21}$ and $\eta$, for $\pi$ pulse scheme in (a) and for LR optimized pulse scheme in (b).}\label{fig8}
		\end{figure}
		From the Hamiltonian parameters $\Omega$ and $\Delta$ in Eq.~(\ref{hp}), we can obtain the relationship between error sensitivity $q$ and $\delta\omega_{21}$ in Fig.~\ref{fig3}(a). {It can be found that LR optimized pulse is better than the $\pi$ pulse in case of $\delta\omega_{21}T<14.76$.} Here, for the  $\pi$ pulse, we have $\theta=\pi t/T$ and $\beta=\pi/2$. Moreover, by
solving numerically the Schrödinger equation with the initial state  $|01\rangle$, a comparison between the 
fidelity of population $|10\rangle$ at final time $T$ based on
the  $\pi$ pulse and invariant-based shortcuts are depicted in Fig.~\ref{fig3}(b). {We can find the fidelity of LR optimized pulse have a broad range above 99\%, demonstrating its robustness against leakage errors.} As shown in Fig.~\ref{fig8}(a) and Fig.~\ref{fig8}(b), we plot the change of fidelity distribution under the variations of r parameters of $\delta\omega_{21}T$ and $\eta$ for the two schemes. It can be found that the invariant-based shortcut technique outperforms its competitors and features a broad
range of high efficiencies. {This provides an optimized and robust method to realize the high fidelity quantum operations in the presence of different experimental parameters.}
	\subsection{Analysis of Leakage Suppression}\label{sec3.3}
		\begin{figure}
		\centering
		\includegraphics[width=7.6cm,height=5.6cm]{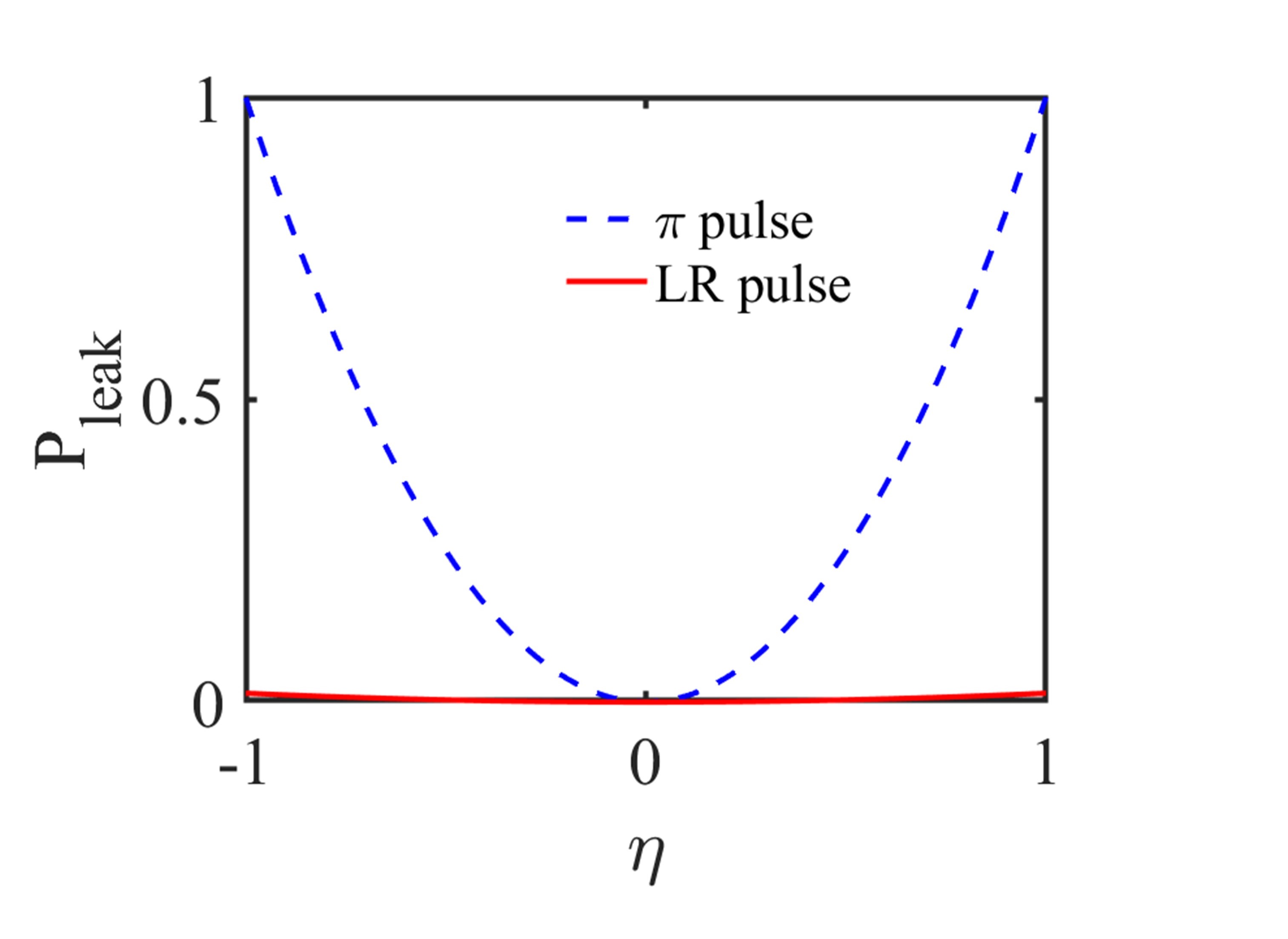}
		\caption{Leakage probability as a function of the relative coupling strength $\eta$ for $\pi$ pulse scheme (blue, dotted line) and LR optimized pulse scheme (red, solid line).}\label{fig6}
	\end{figure}
	
	For leakage probability, it is expressed through the following formula~\cite{4kkq-x5km} as
	\begin{eqnarray}\label{eq37}
		\begin{aligned}
			P_{\rm leak}&=\frac{1}{d_{Q}}\text{Tr}[P_{A}U(T)P_{Q}U(T)^{\dagger}]\\
			&=\frac{1}{d_{Q}}\text{Tr}[P_{A}U_{1}(T)P_{Q}U(T)_{1}^{\dagger}]\\
			&=\sum_{j=1}^{d_{Q}}\sum_{k=1}^{d_{A}}|\langle a_{k}| U_{1}(T)|q_{j}\rangle|^2,
		\end{aligned}
	\end{eqnarray}
	where $P_{Q}$  and $P_{A}$ are the corresponding projectors onto computational subspace spanned by the sates $|q_{j}\rangle$ with $j=1\rightarrow d_{Q}$, and leakage subspace spanned by the sates $|a_{k}\rangle$ with $k=1\rightarrow d_{A}$, {and $d_{Q}$ and $d_{A}$ denote the dimensions of the computational and leakage subspaces, respectively.} 
$P_{Q}$ and $P_{A}$ satisfy the following condition that $P_{Q}+P_{A}=I$. $U(t)$ is full time-evolution operator with $U(t)=U_{0}(t)U_{1}(t)$, where
	\begin{eqnarray}\label{eq38}
		U_0(t)&=\mathcal{T}\exp\left[-i\int_0^tH_0(t^{\prime})dt^{\prime}\right],\\
		U_1(t)&=\mathcal{T}\exp\left[-i\int_0^tH_1(t^{\prime})dt^{\prime}\right].
	\end{eqnarray}
	{$\mathcal{T}$  denotes the time-ordering operator, $U_0(t)$ and $U_1(t)$ are the time evolution operators in the computational subspace with Hamiltonian $H_0$ and leakage subspace with  $H_1= \eta V $ in Eq.~(\ref{eq21}), respectively.} By using time-dependent perturbation theory, the unitary operator $U_1$ can be approximated as
	\begin{equation}
		U_1(T)\approx I-i\int_0^T\eta V(t)dt.
	\end{equation}
	As a result, the leakage probability can be rewritten as
	\begin{equation}
		P_{\rm leak}=\frac{1}{d_Q}\sum_{j=1}^{d_Q}\sum_{k=1}^{d_A}\int_0^Tdt\int_0^Tdt^{\prime}\eta^2\mathcal{A}_{j\to k}(t)\mathcal{A}_{j\to k}^*(t^{\prime}),
	\end{equation}
	where 
	\begin{equation}
		\mathcal{A}_{j\to k}(t)=\langle a_k|V(t)|q_j\rangle.
	\end{equation}
	Here, the Hilbert space basis vectors is defined as
	\begin{eqnarray}\label{eq39}
		\begin{aligned}			
			|q_1\rangle&=|\phi_{0}(t)\rangle\\
			|q_2\rangle&=|\phi_{1}(t)\rangle\\
			|a_1\rangle&=|\phi_{2}(t)\rangle\\
			|a_2\rangle&=|\phi_{3}(t)\rangle.			
		\end{aligned}
	\end{eqnarray}
	So that, we have
	\begin{eqnarray}
		\begin{aligned}	
			\mathcal{A}_{1\to 1}(t)&=\langle a_1|V_1(t)|q_1\rangle = \langle \phi_{2}(t)|V_1(t)||\phi_{0}(t)\rangle  ,\\
			\mathcal{A}_{1\to 2}(t)&=\langle a_2|V_1(t)|q_1\rangle =\langle \phi_{3}(t)|V_1(t)||\phi_{0}(t)\rangle  ,\\
			\mathcal{A}_{2\to 1}(t)&=\langle a_1|V_1(t)|q_2\rangle =\langle \phi_{2}(t)|V_1(t)||\phi_{1}(t)\rangle  ,\\
			\mathcal{A}_{2\to 2}(t)&=\langle a_1|V_1(t)|q_2\rangle =\langle \phi_{3}(t)|V_1(t)||\phi_{1}(t)\rangle . 
		\end{aligned}	
	\end{eqnarray}
	Finally, we can calculate the leakage probability as follows
	\begin{eqnarray}
		\begin{aligned}	
			P_{\rm leak}&=\frac{1}{2}\eta^2(\int_0^Tdta\int_0^Tdt^{\prime}a^*+\int_0^Tdtb\int_0^Tdt^{\prime}b^*\\
			&+\int_0^Tdtc\int_0^Tdt^{\prime}c^*+\int_0^Tdte\int_0^Tdt^{\prime}e^*),
		\end{aligned}	
	\end{eqnarray}
	where  $a=\frac{\Omega}{2}e^{\frac{i}{2}\Gamma_1}cos(\frac{\theta}{2})e^{-i\beta}e^{i\gamma_{+}}$, $b=\frac{\Omega}{2}e^{\frac{i}{2}\Gamma_2}sin(\frac{\theta}{2})e^{i\gamma_{+}}$,
	$c=\frac{\Omega}{2}e^{\frac{i}{2}\Gamma_1}sin(\frac{\theta}{2})e^{i\gamma_{-}}$, and
	$d=-\frac{\Omega}{2}e^{\frac{i}{2}\Gamma_2}cos(\frac{\theta}{2})e^{i\beta}e^{i\gamma_{-}}$.
			
	In Fig.~\ref{fig6}, we draw the curve of the leakage probability with respect to $\eta$ for the $\pi$ pulse and LR optimized pulse schemes. We can see that the leakage probability scales quadratically with $\eta$ in both schemes, as expected from perturbation theory. The probability induced by the leakage error is largely suppressed for the LR optimized pulse scheme,
exhibiting significant superiority beyond the  $\pi$ pulse scheme. The result suggests that the LR optimization control method can actively restrict the population of the quantum state within the computational subspace, so that robust and high-fidelity quantum state control is achieved.
	
	\section{TWO-QUBIT ISWAP GATE AND ENTANGLED STATE GENERATION}\label{sec4}

{Here, we discuss how to realize a two-qubit iSWAP gate and the preparation of remote entangled states by 
 the control of system evolution governed by the Hamiltonian $H_{\text{sub}}$ in the computational subspace.}  However, the evolution of the hybrid system will be affected by relaxation and dephasing of qubit and photon leakage.
To show the robustness of the designed optimized invariant-based shortcut in the presence of such decoherence parameters, we use the Lindblad master equation to simulate the evolution of the quantum system as~\cite{PhysRevA.79.013819}
\begin{equation}\label{eq30}
	\begin{aligned}
		\dot{\rho}&=i[\rho,H_{\text{}}]+\sum_{j=1}^3\left(\gamma_{1,j}^s\mathcal{D}[\sigma_{s,j}^-]\rho+\frac{\gamma_{\phi,j}^s}2\mathcal{D}[\sigma_{s,j}^z]\rho\right.\\
		&+\frac{g_{s,j}^2\kappa}{\Delta_{s,j}^2}\mathcal{D}[\sigma_{s,j}^-]\rho\biggr)+\sum_{k=1}^3\left(\gamma_{1,k}^t\mathcal{D}[\sigma_{t,k}^-]\rho\right.\\
		&+\frac{\gamma_{\phi,k}^t}2\mathcal{D}[\sigma_{t,k}^z]\rho+\frac{g_{t,k}^2\kappa}{\Delta_{t,k}^2}\mathcal{D}[\sigma_{t,k}^-]\rho\biggr),
	\end{aligned}
\end{equation}
{where $H_{\text{}}$ is the control Hamiltonian of hybrid system and $\rho$ is density matrix, Lindblad operator is $\mathcal{D}[\mathcal{O}]\rho=\mathcal{O}\rho\mathcal{O}^\dagger-(\mathcal{O}^\dagger\mathcal{O}\rho+\rho\mathcal{O}^\dagger\mathcal{O})/2$ expressing the dissipative process, and $\gamma_{1}^{s(t)}=1/T_{1}^{s(t)}$ and $\gamma_\phi^{s(t)}=1/T_\phi^{s(t)}$ are relaxation and dephasing rates of qubits, respectively. The rate $g_{s(t)}^{2}\kappa/\Delta_{s(t)}^{2}$ is the Purcell relaxation in the qubits, induced by photon leakage from the resonator.
In this work, all experimentally feasible parameters are chosen as ~\cite{landig2019virtual,PRXQuantum.3.010352,PhysRevA.101.032343}: $g_s/2\pi=90\mathrm{MHz}, g_t/2\pi=200\mathrm{MHz}, T_1^s=20\mathrm{ms}, T_{\phi}^{s}=3\mathrm{\mu s}, T_{1}^{t}=T_{\phi}^{t}=60\mathrm{\mu s}, \kappa/2\pi=0.1\mathrm{MHz}, T=40\mathrm{ns}, \Delta_{s}/2\pi=\Delta_{t}/2\pi=2\mathrm{GHz}$;}
     	\begin{figure}
     	\centering
     	\includegraphics[width=10cm,height=5cm]{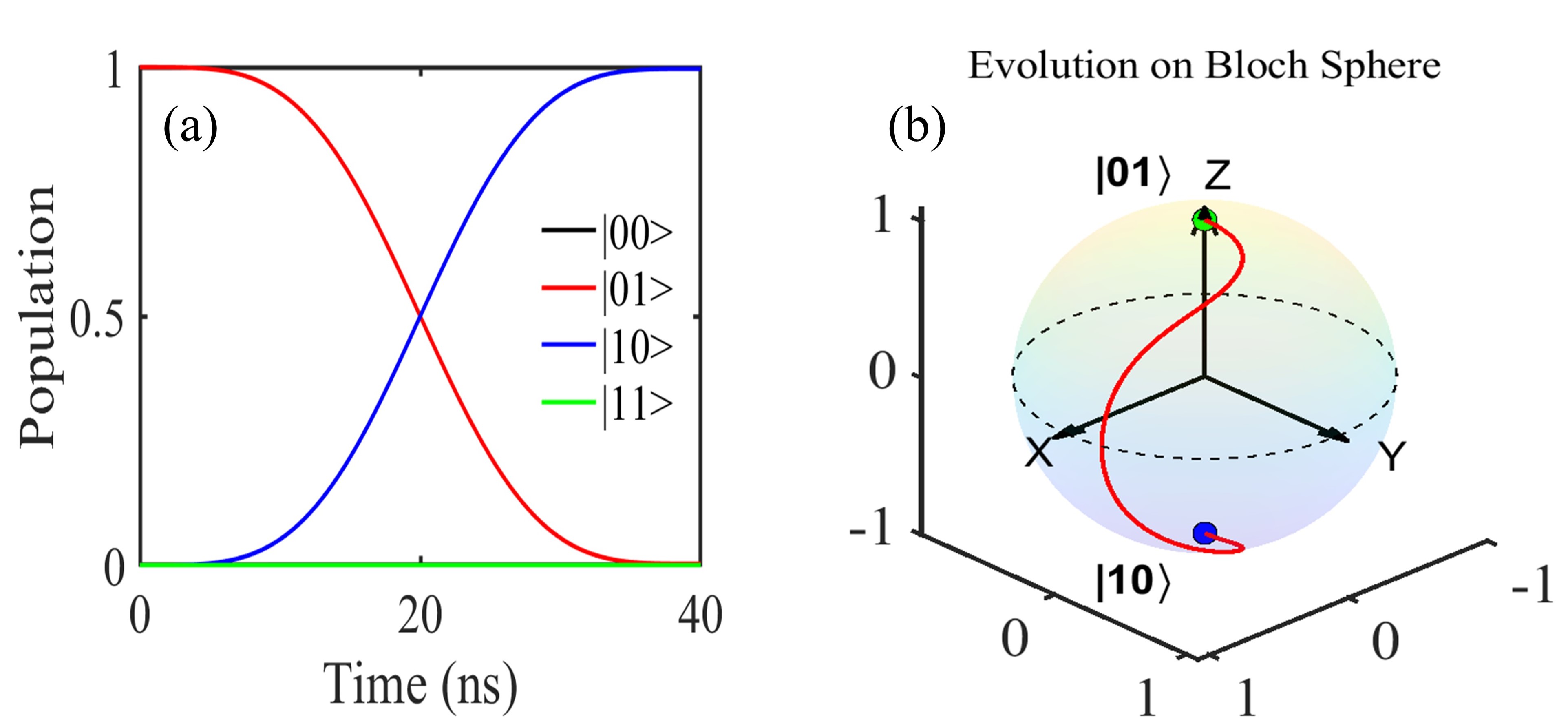}
     	\caption{(a) Time evolution of the state populations of state $|00\rangle$, $|01\rangle$, $|10\rangle$, and $|11\rangle$. (b) The evolutionary path of the hybrid system on Bloch sphere, start (yellow, solid point), end (blue, solid point) and evolutionary path (red, solid line).}\label{fig2}
     \end{figure}

First, we examine the dynamic process of the scheme for implementing the iSWAP gate. In the computational subspace spanned by the basis $\{|00\rangle , |01\rangle , |10\rangle , |11\rangle\}$, 
the control Hamiltonian $H_{\text{sub}}$ is related to the base $|01\rangle$ and $|10\rangle$, and the base $|00\rangle$ and $|11\rangle$ remain unchanged in the whole time evolution of system. In this case, the control Hamiltonian of computational subspace can be rewritten as
\begin{eqnarray}
H_{\text{sub}}'=\underbrace{\begin{pmatrix}-\frac{\Delta}{2}&\frac{\Omega}{2}\\ \frac{\Omega}{2}&\frac{\Delta}{2}\end{pmatrix}}_{H_{\text{sub}}}\oplus{\begin{pmatrix}1&0\\ 0&1\end{pmatrix}}.
	\end{eqnarray}
When the evolution time $T$ satisfies the boundary conditions of Eq.~(\ref{eq40}) and Eq.~(\ref{eq42}). And $\theta$ satisfies Eq.~(\ref{eq29}).
	Then the unitary operation of the system can be expressed  as
	\begin{equation}
		U_{\mathrm{iswap}}=\begin{pmatrix}1&0&0&0\\0&0&i&0\\0&i&0&0\\0&0&0&1\end{pmatrix}.
		\end{equation}
		Obviously, it is the iSWAP gate. To demonstrate the robustness of the iSWAP scheme, we numerically evaluate its performances under actual conditions including decoherence and control errors.
		\begin{figure}
		\centering
		\includegraphics[width=9.6cm,height=4cm]{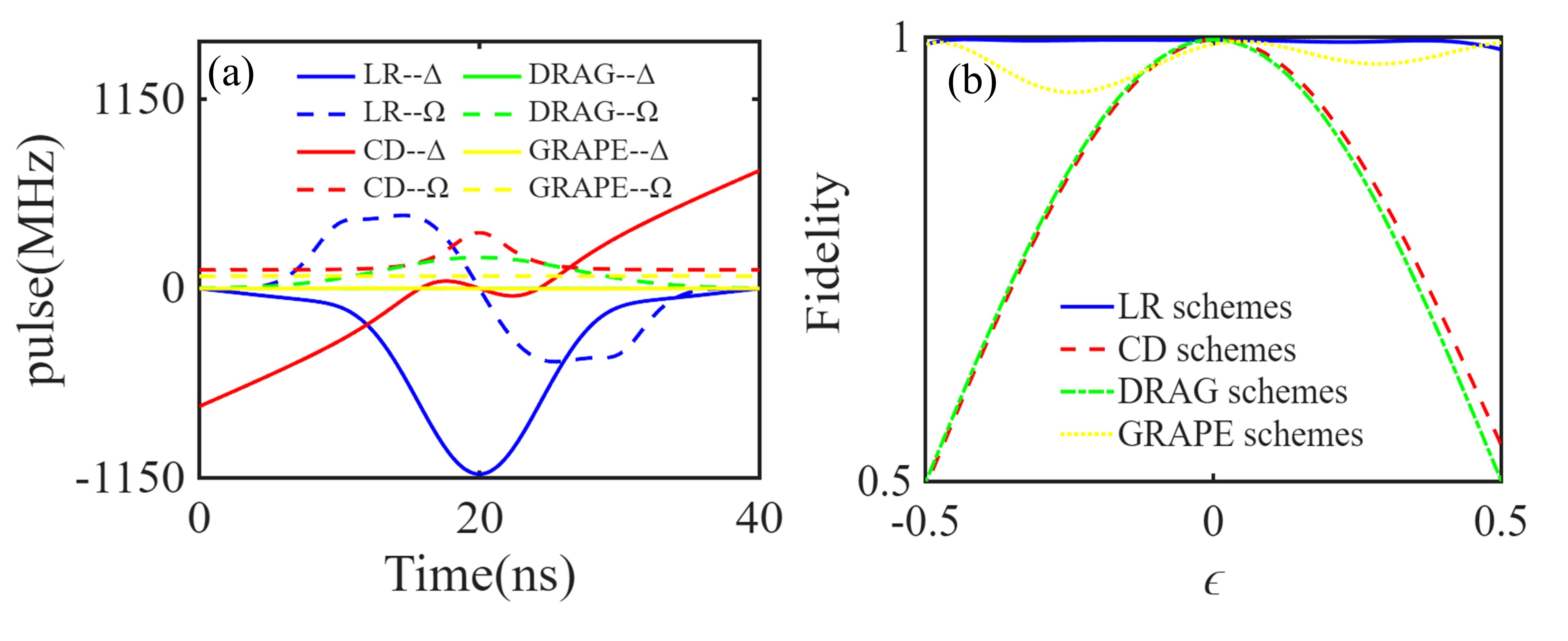}
		\caption{{(a) Pulses that change with respect to time, for LR scheme pulse with $\Delta$ (blue, solid line), $\Omega$ (blue, dotted line), DRAG scheme with $\Delta$ (green, solid line), $\Omega$ (green, dotted line), CD scheme with $\Delta$ (red, solid line), $\Omega$ (red, dotted line), and LBFGS-GRAPE scheme with $\Delta$ (yellow, solid line), $\Omega$ (yellow, dotted line ). (b) Robustness analysis for LR optimized scheme (blue, solid line), DRAG scheme (green dotted line), CD scheme (red dashed line), and LBFGS-GRAPE scheme (yellow dash-dotted line).}}\label{fig4} 
	\end{figure}
	
	As shown in the Fig.~\ref{fig2}(a), we can find that for the prepared initial state $|01\rangle$, its population decreases over time and eventually reaches 0 after time evolution period of a two-qubit iSWAP gate, while the population of state $|10\rangle$ increases over time and almost reaches 1.  Meanwhile, the populations of state $|00\rangle$ and state $|11\rangle$ remain almost vanishing in the whole time evolution. The evolutionary path of the system can be clearly demonstrated on Bloch spheres, as shown in Fig.~\ref{fig2}(b).
	
	Next, we will discuss the robustness of the LR optimized scheme against the control error, in comparison with the CD driving scheme~\cite{PhysRevA.111.012621}, where this work paves the way for  quantum computing based on hybrid spin-superconducting systems and can be extended to other hybrid quantum systems. To realize the above two-qubit iSWAP gate, the Rabi frequency and detuning of the Hamiltonian $H_{\text{sub}}$ in the  CD shortcut scheme can be written as:
	\begin{eqnarray}
		\Delta_{\mathrm{sta}}&=&\Delta-\frac{d}{dt}\left\{\arctan\left[\frac{-\Omega\dot{\Delta}}{\Omega(\Delta^2+\Omega^2)}\right]\right\},\cr\cr
		\Omega_{\mathrm{sta}}&=&\Omega\sqrt{1+\frac{(\Omega\dot{\Delta})^2}{\Omega^2(\Delta^2+\Omega^2)^2}} ,
		\end{eqnarray}
		{where $\Delta$ and $\Omega$ satisfy Landau-Zener pulse protocol~\cite{SHEVCHENKO20101}, with the pulse shape being $\Delta=-\Gamma/2+\Gamma t/T$ and $\Omega=\Omega_0$, where $\Gamma=2\pi\times230\text{MHz}$ and $\Omega_0=2\pi\times18\text{MHz}$. 
		For comparison, we also consider the DRAG pulse scheme and to simulate the fidelity of spin-superconducting entangling gates,  defined in the following form~\cite{PhysRevLett.103.110501,dtdk-kc2b,PhysRevLett.116.230503,PhysRevA.111.012621}:
	\begin{equation}
		\Delta_{\mathrm{DRAG}}(t) =\beta \, \Omega(t)^2,  \quad
		\Omega_{\mathrm{DRAG}}(t) = \Omega_0 \exp \left[ -\frac{(t - T/2)^2}{2\sigma^2} \right], 	
		\end{equation}
		with $\beta=0$, $\Omega_0/2\pi = 30$ MHz, $T = 40$ ns, $\sigma = T/6$.
		{Furthermore, we employ the LBFGS-GRAPE optimal control method~\cite{PhysRevA.84.022305,deFouquieres2011,Khaneja2005} that optimizes the ensemble-averaged fidelity over the error range $\epsilon \in [-0.2, 0.2]$ to further demonstrate the effectiveness of the LR optimized scheme. The algorithm iteratively optimizes the nominal pulses $\Omega_{\text{nom}}(t)$ and $\Delta(t)$ by minimizing the cost function
		\begin{align}
		J = \frac{1}{N_\epsilon}\sum_{i=1}^{N_\epsilon}\left[1 - F(\epsilon_i)\right] + \alpha\sum_i x_i^2,	
		\end{align}
		where $F(\epsilon_i) = |\langle \psi_{\text{target}}|\psi_T(\epsilon_i)\rangle|^2$ is the state fidelity for the $i$-th error sample, and $\alpha$ is a penalty coefficient.  The gradients are obtained via the adjoint-state method with the chain rule, and the pulses are updated using LBFGS quasi-Newton algorithm until convergence. The detailed procedure is summarized in Algorithm 1.}
		
	\begin{algorithm}[htbp]
		\caption{LBFGS-GRAPE optimal pulse optimization}
		\label{alg:grape}
		\begin{algorithmic}[1]
			\Require Initial state $|\psi_0\rangle = |01\rangle$, target state $|\psi_{\text{target}}\rangle = |10\rangle$, 
			time steps $N_t$, total duration $T$, error set $\{\epsilon_i\}_{i=1}^{N_\epsilon} \in [-0.2, 0.2]$,
			penalty $\alpha$, scaling factor $s=2\pi\times1$ MHz, leakage parameter $\eta$
			\Ensure Optimized broadband pulses $\Omega(t)$, $\Delta(t)$, average fidelity $\bar{F}$
			
			\State Initialize dimensionless pulses $\mathbf{x}_0 = [\Omega_0; \Delta_0]$
			\Repeat
			\State Recover nominal pulses: $\Omega_{\text{nom}} = \mathbf{x}_{1:N_t} \cdot s$, $\Delta = \mathbf{x}_{N_t+1:2N_t} \cdot s$
			\State Initialize accumulators: $J_{\text{sum}} = 0$, $\nabla J_{\text{sum}} = 0$
			
			\For{each error sample $\epsilon_i \in [-0.2, 0.2]$}
			\State Apply perturbed fields: 
			$\Omega_{\text{main}} = (1+\epsilon_i)\Omega_{\text{nom}}$, 
			$\Omega_{\text{leak}} = (\eta+\epsilon_i)\Omega_{\text{nom}}$
			\State Construct $H(t)$ from $\Omega_{\text{main}}(t)$, $\Omega_{\text{leak}}(t)$, $\Delta(t)$
			\State Forward propagate: $\psi_{k+1} = e^{-iH_k dt}\psi_k$, $k=1,\dots,N_t$
			\State Compute fidelity: $F_i = |\langle \psi_{\text{target}}|\psi_T\rangle|^2$
			\State Accumulate cost: $J_{\text{sum}} \leftarrow J_{\text{sum}} + (1 - F_i)$
			\State Compute gradients $\partial F_i/\partial \Omega_{\text{nom}}$ via adjoint-state method with chain rule:
			$\displaystyle \frac{\partial J_i}{\partial \Omega_{\text{nom}}} = -(1+\epsilon_i)\frac{\partial F_i}{\partial \Omega_{\text{main}}} - (\eta+\epsilon_i)\frac{\partial F_i}{\partial \Omega_{\text{leak}}}$
			\State Accumulate gradients: $\nabla J_{\text{sum}} \leftarrow \nabla J_{\text{sum}} + \nabla J_i$
			\EndFor
			
			\State Ensemble average: 
			$J = J_{\text{sum}}/N_\epsilon + \alpha\sum_i x_i^2$,
			$\nabla J = \nabla J_{\text{sum}}/N_\epsilon + 2\alpha \mathbf{x}$
			\State Update $\mathbf{x}$ using LBFGS quasi-Newton update
			
			\Until{$\|\nabla J\| < \text{tol}$ or max iterations reached}
			\State \Return $\Omega(t) = \Omega_{\text{nom}}(t) \cdot s$, $\Delta(t)$, $\bar{F} = \langle F_i \rangle$
		\end{algorithmic}
	\end{algorithm}
		
		{In Fig.~\ref{fig4}(a), we show the corresponding  the pulses of Rabi frequencies and detunings in the LR optimized, CD, DRAG and LBFGS-GRAPE protocols, respectively.}
Here the parameters in LR optimized pulse scheme can be calculated through the Eqs.~(\ref{eq16}), (\ref{eq28}), and (\ref{eq29}). 	
	In the following, we consider control error of Rabi frequency in the  $H_{\text{sub}}$ as a  perturbation term. Generally, the perturbing Hamiltonian is assumed to take the form
	 \begin{eqnarray}
	 	H_{\text{err}}=\frac{1}{2}{\begin{pmatrix}
	 			0& \epsilon\Omega\\
	 			\epsilon\Omega&0
 			\end{pmatrix}}.
	 \end{eqnarray}	 }
	 Then, the whole Hamiltonian to control the time evolution of the system is  $H_{\text{sub-err}}=H_{\text{sub}}+H_{\text{err}}$, that is,
	   \begin{equation}\label{eq33}
	 	\begin{aligned}
	 		H_{\text{sub-err}} &= \frac{\Delta}{2}\left(|10\rangle\langle 10| - |01\rangle\langle 01|\right) \\
	 		&+ \frac{(1+\epsilon)\Omega}{2}\left(|01\rangle\langle 10| + |10\rangle\langle 01|\right) .
	 	\end{aligned}
	 \end{equation}	 
	Substituting  $H_{\text{sub-err}}$ into Eq.~(\ref{eq30}), we can obtain the actual evolved density matrix $\rho_{\text{re}}$. {When substituting it into $F_{\mathrm{tr}}=\mathrm{Tr}(\sqrt{\sqrt{\rho_{\mathrm{id}}}\rho_{\mathrm{re}}\sqrt{\rho_{\mathrm{id}}}})^{2}$ to get the final fidelity in the presence of decoherence and control error, where $\rho_{\text{id}}$ denotes the ideal density matrix after the time evolution.} We plot the fidelities with respect to the error parameter $\epsilon$ in the Fig.~\ref{fig4}(b). 
	 It can be found that the LR optimized scheme maintains high fidelity even under substantial pulse perturbation, exhibiting its robustness against control errors. {The DRAG and CD schemes exhibit comparable performance. Their fidelities overlap and are generally lower than those of the LR and LBFGS-GRAPE schemes over all control error ranges.}  These results indicate that the LR optimized technique offers improved robustness over the other methods in the presence of control errors.
	{When considering the existence of leakage error, we discuss the robustness of the scheme against control error. For the case where both leakage error and control error exist and with $\delta\omega_{21}=0$, our control Hamiltonian can be written as:}
	\begin{equation}\label{eq50}
		\begin{aligned}
			H_{\text{sub-leak-err}} &= \frac{\Delta}{2}\left(|10\rangle\langle 10| - |01\rangle\langle 01|\right) {+\frac{\Delta}{2}\left(|20\rangle\langle 20| - |02\rangle\langle 02|\right)}  \\
			&+ \frac{(1+\epsilon)\Omega}{2}\left(|01\rangle\langle 10| + |10\rangle\langle 01|\right) \\
			& + \frac{(\eta+\epsilon) \Omega}{2}\left(|01\rangle\langle 02| + |02\rangle\langle 01|\right) \\
			& + \frac{(\eta+\epsilon) \Omega}{2}\left(|10\rangle\langle 20| + |20\rangle\langle 10|\right)  .
		\end{aligned}
	\end{equation}	
		\begin{figure}[htbp]
		\centering
		\includegraphics[width=6.6cm,height=5cm]{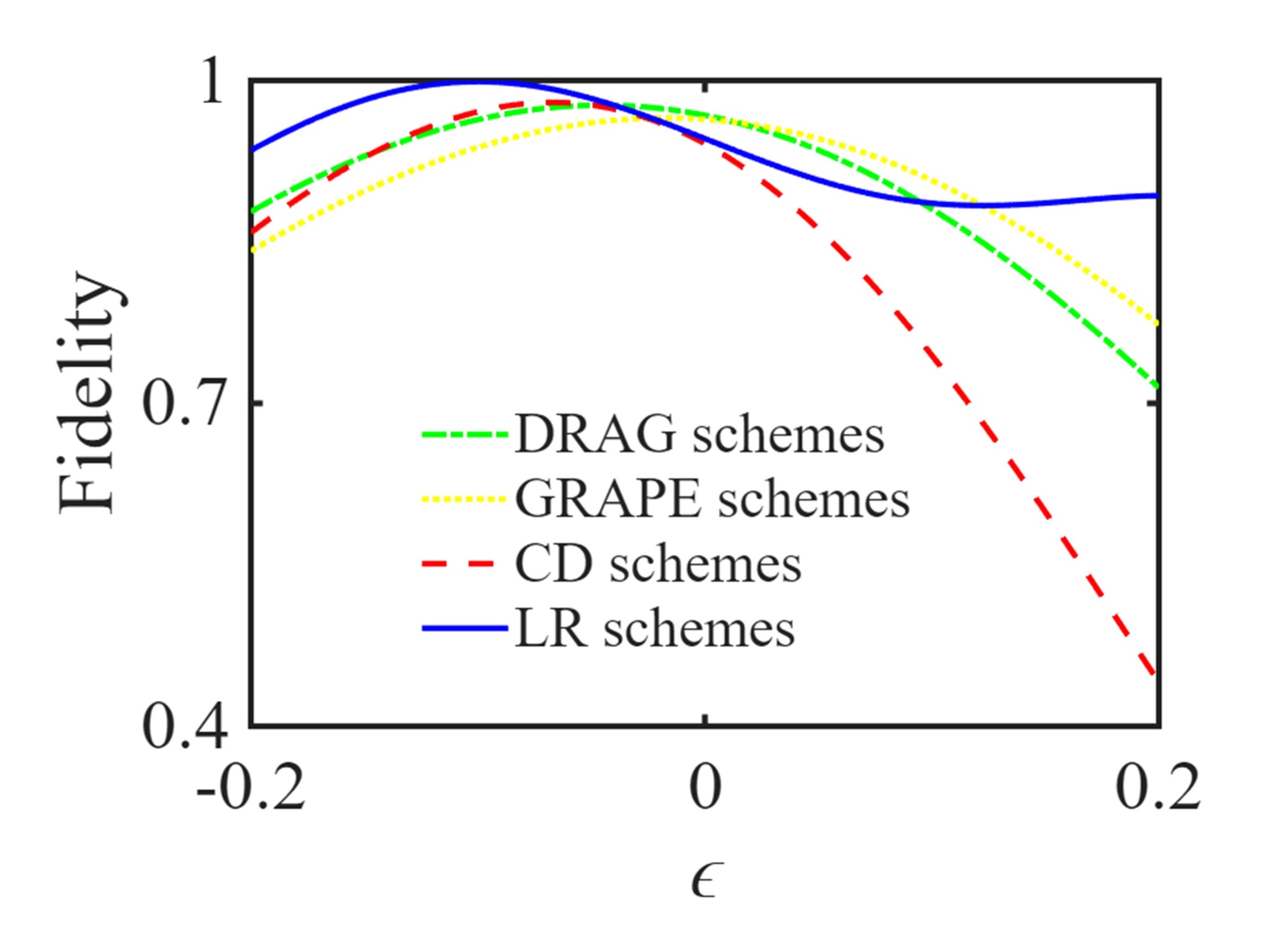}
		\caption{ {Conduct a robustness analysis of the LR optimization scheme (blue solid line), the DRAG scheme (Green dotted line), the CD scheme (red dashed line), and the LBFGS-GRAPE scheme (yellow dash-dotted line) when considering the presence of leakage errors with consider $\eta = 0.1$.}}\label{fig9} 
	\end{figure}
	 
	{As shown in Fig.~\ref{fig9}, we plot the fidelity as a function of the error parameter $\epsilon$, with $\eta = 0.1$. }
	It can be observed that the scheme based on LR maintains high fidelity within a wide range of control errors even in the presence of leakage errors, demonstrating its strong robustness. The DRAG scheme also exhibits good robustness, while its fidelity is lower than that of the LR scheme in most regions of $\epsilon$. {The LR, DRAG, and LBFGS-GRAPE schemes exhibit high fidelity in the presence of leakage errors, although their precedence varies across different $\epsilon$ intervals. By contrast, the CD scheme always delivers the lowest fidelity.} These results indicate that, in the presence of leakage and decoherence, the LR optimization method offers advantages over the DRAG, CD, and LBFGS-GRAPE methods in terms of robustness against control errors.
		{Furthermore, we apply the LR optimized method to the preparation of Bell entangled states.} By simply adjusting the boundary conditions of the parameter $\theta$ of the Eq.~(\ref{eq40}), the system can evolve from a separable initial state to an entangled state.
		To control this hybrid system to prepare entangled state in the computational subspace, the evolutionary operator $U$ of system can be expressed as
	\begin{equation}\label{eq35}
		U=\sum_{n} \vert \Psi_{n}(T) \rangle \langle \Psi_{n}(t) \vert,
	\end{equation}
	with $\vert \Psi_{n}(T) \rangle=e^{i\gamma_n}|\phi_n(t)\rangle$ from Eq.~(\ref{eq17}). As a result, we can obtain the time evolution operator $U$ as
	\begin{eqnarray}\label{eq36}
		U={\begin{pmatrix}
				a& b\\
				-b^*&a^*
		\end{pmatrix}},
	\end{eqnarray}
	where $a=e^{i\gamma}cos^2(\theta/2)+e^{-i\gamma}sin^2(\theta/2)$, $b=isin(\gamma)sin(\theta)e^{-i\beta}$.
	By adjusting the function of $\theta$, we can  prepare arbitrary two qubit entangled states.
	
	For example, to prepare a Bell states, the parameter $\theta$, by adjusting the corresponding boundary conditions, can be designed as 
	    \begin{equation}\label{eq34}
		\theta=\frac{\pi t}{2T}.
	\end{equation}

\begin{figure}[htbp]
		\centering
		\includegraphics[width=9.6cm,height=8cm]{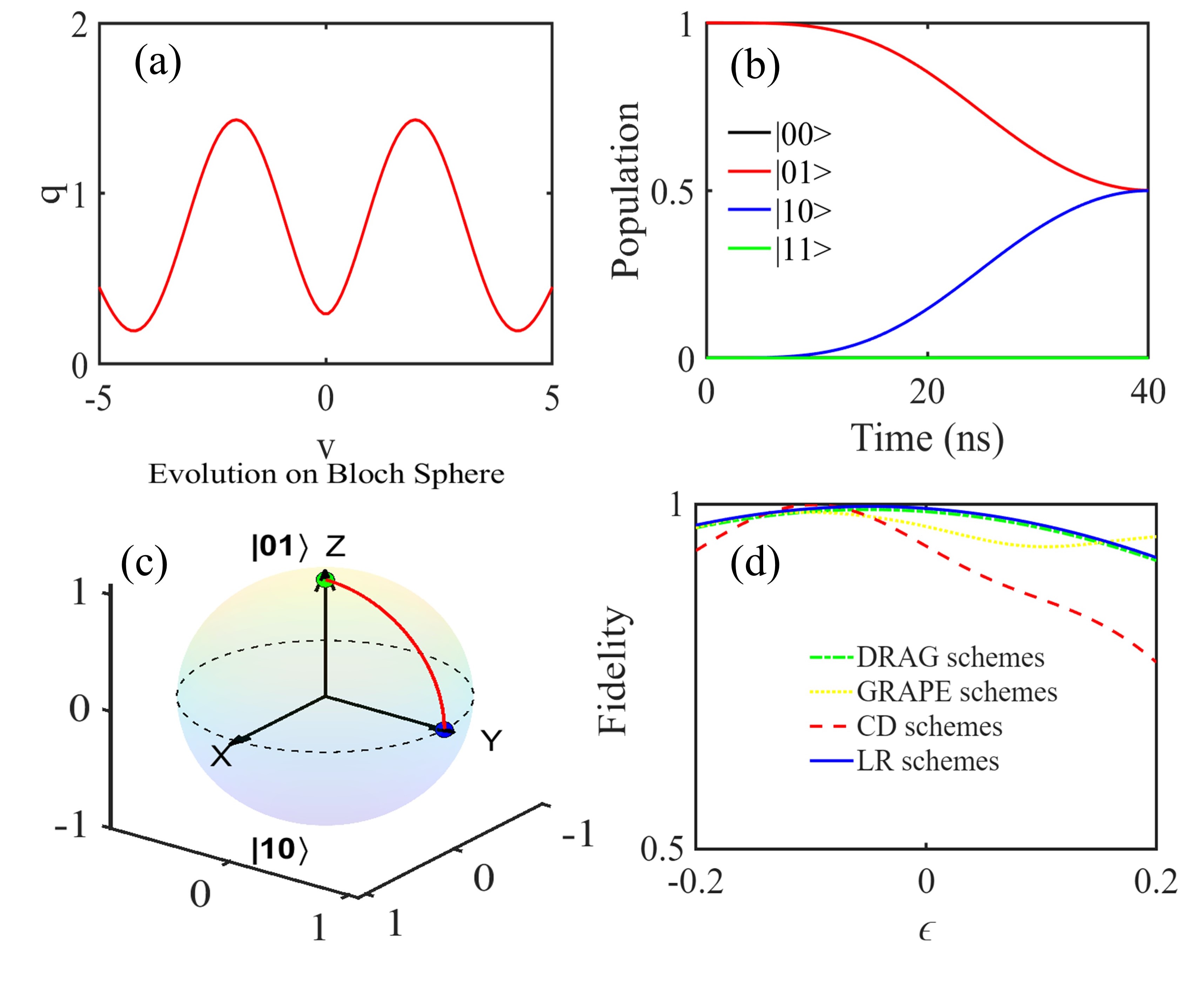}
		\caption{ (a) Leakage error sensitivity q in Eq.~(\ref{eq26}). (b)  Time evolution of the state populations of state $|00\rangle$, $|01\rangle$, $|10\rangle$, and $|11\rangle$. (c) The evolutionary path of the hybrid system on Bloch sphere, start (yellow, solid point), end (blue, solid point) and evolutionary path (red, solid line). (d)  {Conduct a robustness analysis of the LR optimization scheme (blue solid line), the DRAG scheme (Green dotted line), the CD scheme (red dashed line), and the LBFGS-GRAPE scheme (yellow dash-dotted line) when considering the presence of leakage errors with consider $\eta$ = 0.1 when considering the presence of leakage errors.}}\label{fig5}
	\end{figure}

	From Eq.~(\ref{eq26}) and Eq.~(\ref{eq28}), the relationship between error sensitivity $q$ and $\nu$ can be obtained, which is shown in Fig.~\ref{fig5}(a).	
	To minimize the $q$, we select the  simplest $\pi$ pulse can be achieved by setting $v=0$, that  $\beta=\frac{\pi}{2}$.  By solving the Lindblad master equation in Eq.~(\ref{eq30}), we can get the populations of state $|00\rangle$, $|01\rangle$, $|10\rangle$, and $|11\rangle$ under decoherence, respectively, plotted in  Fig~\ref{fig5}(b). {We can see
that the populations of state $|01\rangle$ and $|10\rangle$ are approximately 0.5, while the populations of state $|00\rangle$ and $|11\rangle$ remains zero, suggesting that a two-qubit Bell state is generated.}
 Its evolutionary path is shown in Fig.~\ref{fig5}(c). {Similarly, in the presence of leakage errors and decoherence, considering the presence of control error, the fidelity of the Bell state with $\epsilon$ is shown in Fig.~\ref{fig5}(d). The LR scheme maintains high fidelity above 98\% over a broad range of $\epsilon$, revealing its strong robustness against control errors. {These results indicate that, even in the presence of leakage and decoherence, the LR optimized method always show its ultrahigh robustness against	control errors in the preparation of entangled states.}
The designed LR optimized protocol possesses the property of ``error tolerance" and ``short duration and high fidelity" in the presence of experimental environments makes it have many potential applications in quantum communication and computing.

	In a word, the numerical simulation presented in this section provides comprehensive evidence of the importance for optimizing the performances of the control scheme. By using the optimized invariant-based shortcut, we successfully realize high-fidelity iSWAP gates and Bell state generation under the effect of real decoherence and control error. Most importantly, compared with the CD shortcut, it demonstrates superior robustness against control error, and provides a flexible framework to achieve various quantum operations. These characteristics make the invariant-based shortcut a very promising tool for precisely controlling hybrid quantum systems.

	\section{CONCLUSION}\label{sec5} 
{In conclusion, we have proposed an analytically optimal method to suppress the transducer leakage errors in hybrid spin-superconducting quantum systems.} By using the optimized invariant-based inverse engineering, we have realized an efficient suppression of the transducer leakage errors at a lowest value with $0.01$ by actively minimizing transitions from computational subspace to non-computational subspace. Furthermore, we have derived analytically optimized control pulses that achieve a high-fidelity iSWAP gate and the preparation of remote entangled states in the computational subspace. {Our LR optimized scheme demonstrates a remarkable fidelity exceeding 99\% of the two above quantum operations under realistic decoherence and noise, significantly outperforming conventional $\pi$-pulse , CD-driving, DRAG protocols and  LBFGS-GRAPE methods in terms of robustness.} When considering the situation where there is a leakage error in reality, our scheme still possesses robustness against control error over a wide range.  The primary significance of this work can be summarized in two aspects: Theoretically, it proposes an effective control  solution to the leakage problem caused by the multi-level structure of the hybrid spin-superconducting quantum system; practically, the demonstrated applications to iSWAP gate operation and entanglement preparation in the  hybrid spin-superconducting quantum circuits confirm the immediate utility of the scheme in various quantum information processing tasks. 
The results provide an effective tool for suppressing the leakage errors in high-dimensional hybrid quantum systems, revealing the further applications of optimal control theory in robust quantum computing and communication.

\section*{acknowledgements}

	The authors acknowledge the financial support by National Natural Science Foundation of China~(Grants No.~62471001, No.~12475009, No.~12575032, No.~12075001, and No.~12175001), Innovation Program for Quantum
Science and Technology ~(Grant No. 2582021ZD0301704), Natural Science Research Project in Universities of Anhui Province (Grant No. 2024AH050068), Anhui Provincial Key Research and Development Plan (Grant No. 2022b13020004), Anhui Province Science and Technology Innovation Project (Grant No. 202423r06050004),
A major science and technology project of Henan Province under (Grant No. 221100210400).

		\bibliography{REV}
\end{document}